\documentclass[11pt]{article}
\usepackage{epsfig}
\usepackage{graphicx}
\usepackage{epsfig}

\begin{document}

\title{\bf  First-principles calculations of step formation energies and step interactions on TiN(001)
}
\author{C. V. Ciobanu$^{1,2}$\footnote{Corresponding author}, D. T. Tambe$^1$ and V. B. Shenoy$^1$ \\
$^1$Division of Engineering, Brown University, Providence, Rhode
Island 02912, USA \\ $^2$Division of Engineering, Colorado School
of Mines, Golden, Colorado 80401, USA}
\date{\today}
\maketitle

\begin{abstract}
We study the formation energies and repulsive interactions of
monatomic steps on the TiN(001) surface, using  density functional
total-energy calculations. The calculated formation energy of
[100] oriented steps agree well with recently reported
experimental values; these steps are shown to have a rumpled
structure, with the Ti atoms undergoing larger displacements than
the N atoms. For steps that are parallel to [110], our
calculations predict a nitrogen (N) termination, as the
corresponding formation energy is several hundred meV/\AA \
smaller than that of Ti-terminated steps.
\end{abstract}

\maketitle \newpage

%
% Introduction:
%

Titanium nitride thin films have attracted sustained interest for
diverse technological application partly because of fast diffusion
characteristics on the (001) surfaces. Although there have been a
large number of experimental reports on different aspects related
to thin film growth on TiN surfaces,  few atomistic studies have
been performed so far. Recognizing the need for accurate atomic
scale properties, several different groups have recently employed
density functional total-energy methods to study the structure and
energies of low index TiN surfaces \cite{Kobayashi,
ukTiN-2000,uiucTiN-2003}, as well as the diffusion barriers on
these surfaces \cite{uiucTiN-2003}. To our knowledge, similar
studies for the stepped surfaces have not yet been attempted.
Motivated by recent experimental work
\cite{K-SS-2002,K-PRL-2002decay, K-PRL-2002,K-PRB-2003} that
addresses the determination of absolute step formation energy and
stiffness from equilibrium shape fluctuations and decay of
two-dimensional islands on TiN surfaces, we study here the
structure and energetics of monatomic steps on TiN(001). While the
[001] step edge is made of alternating Ti and N atoms, the nature
(i.e. N-terminated or Ti-terminated) of the [110] oriented steps
has not been elucidated. Calculations of step formation energies
for the two types of edge terminations can give insight into the
structure of the step. In this paper, we report the formation
energies as well as the strength of repulsive interactions between
the steps, and provide a description of the step structures in
terms of atomic displacements. Our calculations on step energetics
are consistent with the experimental results for [100] steps, and
suggest that the steps along [110] direction are N-terminated.

%
% Density functional calculations: details
%

The density functional calculations were carried out with the VASP
package \cite{Vasp}, using ultra-soft pseudopotentials and the
Perdew-Wang functional form for the exchange-correlation energy
\cite{pw91}. In all the computations we use the experimental value
for the lattice constant of TiN, $a=4.24$\AA; for this lattice
constant, bulk calculations with an eight-atom unit cell and 35
$k$ points yielded a bulk energy per pair of $e_b=-19.2747$ eV.
For surface calculations, the Brillouin zone was sampled using a
$\Gamma$-centered $8\times 8 \times 1$ grid. This sampling yielded
15 $k$ points for the TiN(001) surface, and 21--25 $k$ points for
the stepped surfaces. The ions were relaxed via a
conjugate-gradient algorithm until the total energy converged to
less than 0.001 eV. The energy cutoff for the plane waves was set
to 250eV (18.37 Ry) throughout this study.
%
% flat surfaces : calibration, comparison with recent previous work at UIUC, UK, Jap
%
Before discussing the stepped surfaces, we compare the results for
the flat TiN(001) surface obtained using the above model
parameters with other recent theoretical studies \cite{Kobayashi,
ukTiN-2000,uiucTiN-2003}. It is well known (see, e.g., Ref.
\cite{Kobayashi}) that this surface exhibits a so-called rumpling
reconstruction, where the N atoms relax outwards the surface and
the Ti atoms are pulled inwards. Although somewhat different
computational parameters are used in Refs. \cite{Kobayashi,
ukTiN-2000,uiucTiN-2003} than in the present study, our results
for the displacements of the first and second surface layer agree
with previously reported values (refer to Table~\ref{flatdispl}).
In terms of surface energy of TiN(001), we obtained
$\gamma_{(001)}$=80.6 meV/\AA$^2$, which is also very close to the
value of 81 meV/\AA$^2$ given in Refs. \cite{ukTiN-2000,
uiucTiN-2003}.

\begin{figure}
\begin{center}
\epsfig{file=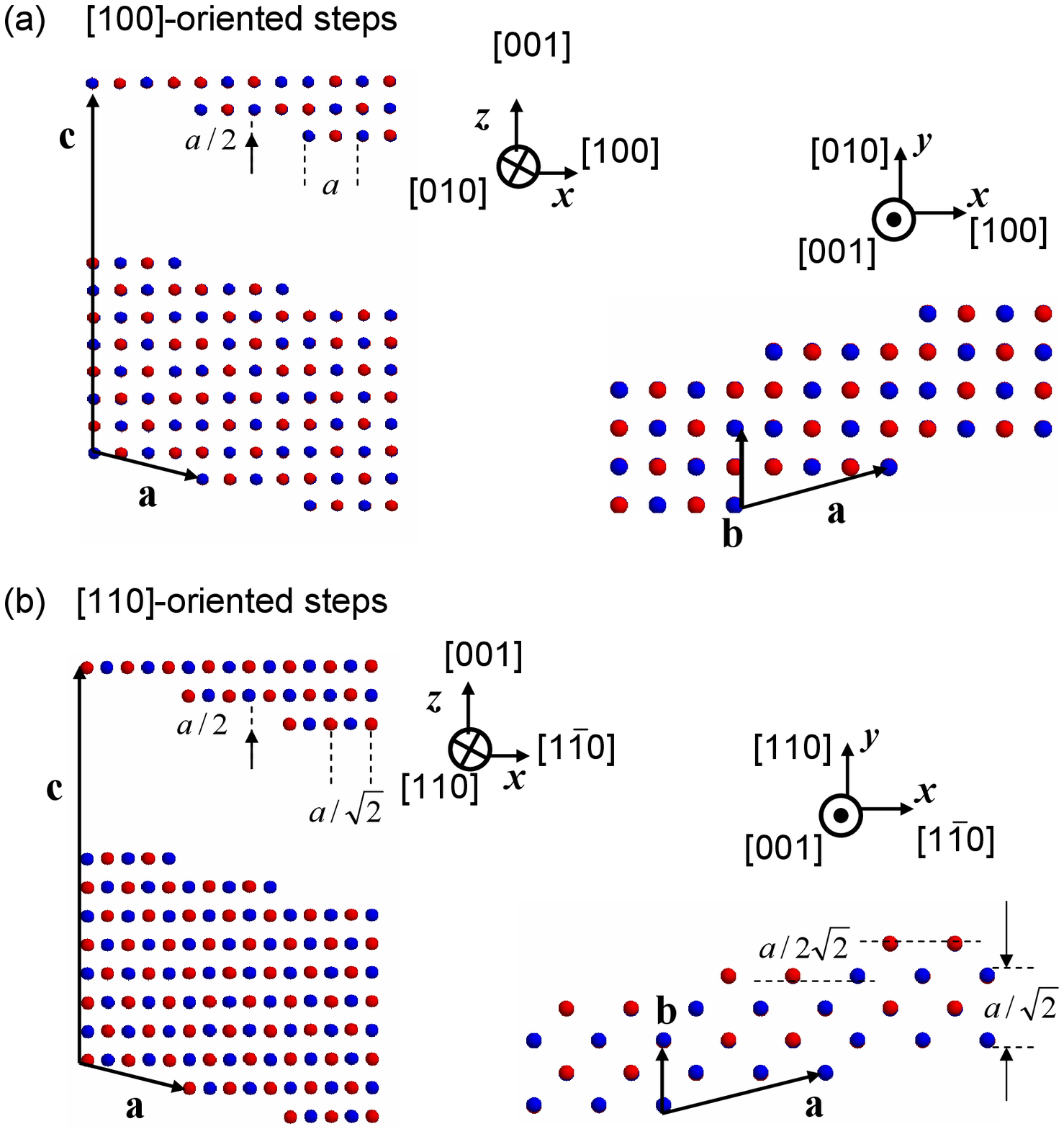,width=12.0cm} %
\caption{Typical configurations of stepped TiN(001) surfaces, with
(a) steps along [100], and (b) steps oriented along [110]. The
periodic vectors of the computational cell are indicated by {\bf
a}, {\bf b}, and {\bf c}. The steps oriented in the [100]
direction are made of alternating N (blue) and Ti (red) atoms as
depicted in (a), while the steps parallel to [110] are either
N-terminated or Ti-terminated  (upper and lower slab surface,
respectively) as shown in fig. (b). }\label{geom}
\end{center}
\end{figure}

%
% stepped surface geometries and boundary conditions
%
In this work we are considering two types of steps on the TiN(001)
surface, which correspond to the [100] and [110] orientations. The
[100] step edge consists of alternating N and Ti atoms along the
step, while the [110] steps have edges that are made of only one
type of atoms, either N or Ti. Fig.~\ref{geom} illustrates the
step structures considered here, and also shows the periodic
supercells used in the density functional calculations. The
dimensions of the supercell are $L_x \times L_y \times L_z$, where
$L_x$ and $L_z$ denote the terrace width and the height of the
supercell, respectively. The dimension $L_y$ is taken equal to the
spatial period in the direction parallel to the step, which is $a$
for the [100] steps and $a\sqrt{2}/2$ in the case of [110] steps.
To create the steps, we employ shifted boundary conditions
\cite{Poon}, in which the amount of shift in the $z$- direction is
the step height ($a/2$) and the shift in the $y$- direction is
determined from requirements of periodicity. Most calculations on
vicinal surfaces were carried out using a slab of TiN with 8
atomic layers (16.96 \AA) and a vacuum thickness of 12\AA.

%
% Results for [100] oriented steps: formation, repulsion, displacements of the atoms at the steps
%

%
% Results for [110] oriented steps: formation (only as average), repulsion (again-ave or order of magnitude stuff)
%  N -terminated, Ti-terminated, when they come together they form the (111) surface -> prety cool .
%  displacements

Similar to our recent work \cite{tacpaper}, the energetics of
steps is studied starting from the ledge energy, defined as
\cite{Poon}
\begin{equation}
\lambda = ( E - N_p e_b-2\gamma_{(001)}A )/2L_y \ ,
\label{lambda-stoic}
\end{equation}
where $E$ is the relaxed total energy of the $N_{p}$  Ti-N pairs
in the supercell, and $A$ is the projected area of the slab on the
(001) plane. The ledge energy is the energy per unit length (along
the step) of the vicinal surface in excess of the surface energy
of the terraces separating the steps: in the case of [100] steps
this excess energy can be defined by Eq.~(\ref{lambda-stoic})
because the system is stoichiometric irrespective of the slab
thickness. For the [110] steps, the only way to preserve
stoichiometry is to have an N-terminated step on one face of the
slab and a Ti-terminated step on the other face (even number of
layers), case in which Eq.~(\ref{lambda-stoic}) will give the
average ledge energy between the two types of [110] steps. On the
other hand, when the number of (001) layers in the computational
cell is odd, the same kind of termination for the [110] steps will
be present on both sides of the slab: in this non-stoichiometric
situation, the ledge energy can  be written as
\begin{equation}
  \lambda=(E-N_N \mu_N -N_{Ti}\mu_{Ti}-2\gamma_{(001)}A)/2L_y,
\label{lambda-nonstoic}
\end{equation}
where $N_N$ ($N_{Ti}$) and $\mu_{N}$ ($\mu_{Ti}$) are the number
of N (Ti) atoms and their chemical potential, respectively. While
the two chemical potentials add up to the bulk cohesion energy per
pair ($\mu_N+\mu_{Ti}=e_b$), the chemical potential of (e.g.) the
N atoms is not known, which leads to ambiguities in the values of
the formation energies of each of the two types of [110] steps. We
will continue with the analysis of ledge energies, but return to
this issue in later paragraphs.

\begin{figure}
\begin{center}
\epsfig{file=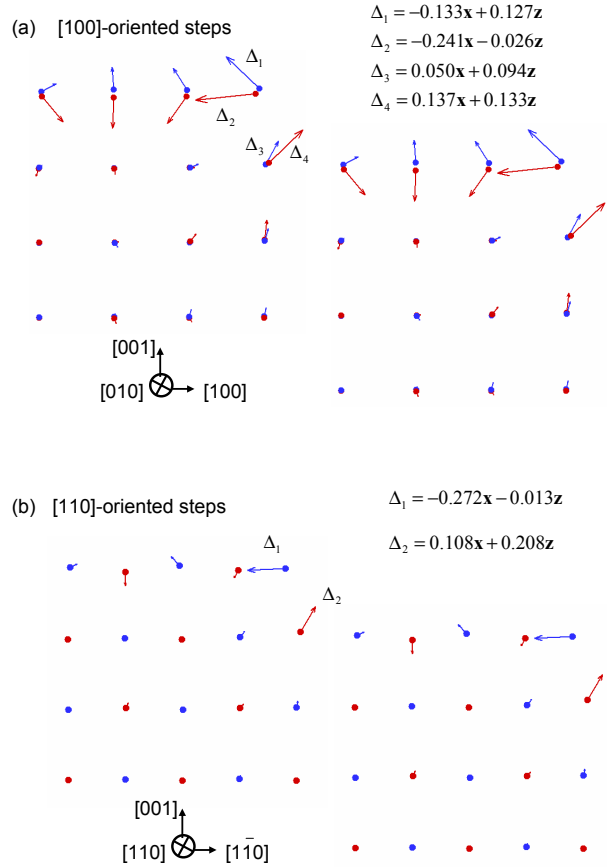,width=10.0cm} %
\caption{(a) Structure of the single-height, [100]-oriented steps
that form a TiN(104) surface. (b) Structure of the N-terminated,
[110]-steps that form a TiN(115) surface. The displacements of N
and Ti atoms are represented by blue and red arrows
(respectively), which lie in the $xz$ plane and are magnified for
clarity. For selected atoms around the step edge the components of
the displacement vectors $\Delta_i (i=1,...,4)$ are given in \AA
ngstrom. }\label{displacements}
\end{center}
\end{figure}

The ledge energy ---as calculated by either
Eq.~(\ref{lambda-stoic}) or (\ref{lambda-nonstoic}), includes both
the step formation and interaction energies. The interaction
between the steps is caused by electrostatic and by elastic
contributions, with comparable magnitudes \cite{tacpaper}. Using
the fact that both the elastic and electrostatic effects give rise
to dipolar interactions \cite{marchenko,Jay}, the ledge energy can
be expressed as
\begin{equation}
    \lambda = \Lambda + \frac{\pi^2}{6}\frac{G}{L_x^2} \ \ \ ,
    \label{lambdalin}
\end{equation}
where $\Lambda$ is the formation energy of a step and $G$ is the
strength of the repulsive interaction between two isolated steps.
The factor $\pi^2/6$ arises since we treat a periodic array of
steps, rather than two individual ones. Similar to Ref.
\cite{tacpaper}, we compute the fitting parameters $\Lambda$ and
$G$ in Eq.~(\ref{lambdalin}) and give the results for the
stoichiometric case in Table \ref{beta-and-gamma}. For the
[100]-oriented steps we found a formation energy of 238 meV/\AA \
, which is in very good agreement with the value of $250\pm 50$
meV/\AA \  obtained from two-dimensional equilibrium island shape
and coarsening measurements on epitaxial Ti(001) layers
\cite{K-SS-2002}. The interaction strength $G$, which has not been
reported so far, is found here to be 548 meV\AA; the result is
similar to the repulsive strength (607 meV/\AA) recently
determined for single-height steps on TaC(001) \cite{tacpaper}.
The structure of the step edges along [100] (shown in
Fig.~\ref{displacements}(a)) is also similar to the one reported
for steps on TaC(001) \cite{tacpaper}, with the metal atoms
undergoing larger displacements than the N atoms.

While it is notable to have found that the calculated formation
energy for [100] steps agrees with experiments \cite{K-SS-2002},
this agreement ceases in the case of [110] oriented steps. The
discrepancy is not surprising for the following two reasons: (a)
in the case of [110] steps, the value we report in Table
\ref{beta-and-gamma} is the arithmetic mean between the formation
energies of N- and Ti-terminated steps and (b) in experiments, it
is not known what type atoms lies at the step edges, and,
furthermore, it is in unlikely that the two terminations occur
with equal probability. Given this situation, we have set out to
investigate further the [110] steps in order to estimate the
individual formation energies of the N- and Ti-terminated steps.
To this end, we have repeated the calculations using supercells
with odd numbers of layers, for which steps are terminated in the
same way on both faces of the slab.

Although we do not consider the smallest possible step separations
when fitting the ledge energies to Eq.~(\ref{lambdalin}), it is
worth noting that for those separations ($L_x=a\sqrt{2}/4$) the
two faces of the slab assume the (111) facet orientations. For the
N- and Ti-terminated (111) facets, the surface energies have been
reported recently by Gall and coworkers \cite{uiucTiN-2003}, which
we will now compare with our own results in the limit of smallest
terrace width. Recalling that these surface energies depend on the
value chosen for chemical potential of either N or Ti (but not
both), we start by choosing a value of $\mu_N$ that exactly
reproduces the value of 85 meV/\AA$^2$ given in Ref.
\cite{uiucTiN-2003}. Under this condition ($\mu_N=-7.49$ eV), the
surface energy that we computed for the Ti-terminated (111) was
373 meV/\AA$^2$, an 8\% difference from the value of $346$
meV/\AA$^2$ given in \cite{uiucTiN-2003}. This difference is
perhaps due to the larger number of layers (thirteen) used in Ref.
\cite{uiucTiN-2003}, which in principle allows for a better
relaxation.\footnote{Other contributing factors could be
differences in the vacuum size, the kinetic energy cut-off for the
plane waves, and the $k$- space sampling} Since we are studying
supercells with different terrace widths (different sizes in the
$x$- direction), in order to keep calculations tractable with our
current computational resources we used a maximum of 9 layers.
While the convergence of the surface energy for (111) facets might
not have been fully reached for 9 layers, the results reported
here are consistent: the average of $\gamma_{N-(111)}$=85
meV/\AA$^2$ and $\gamma_{Ti-(111)}$=373 meV/\AA$^2$ (which have
been obtained from independent odd-layer calculations) differs by
only 1.6 meV/\AA$^2$ from the result of a separate, even-layer
calculation which gives $\overline{\gamma}_{(111)}$=227.4
meV/\AA$^2$ (Table \ref{beta-and-gamma}).

With this preamble on the surface energies of (111) facets, the
formation energies obtained for the N- and Ti- terminated [110]
steps are -14 meV/\AA$^2$ and 846 meV/\AA$^2$, respectively. These
values indicate that it is much more energetically favorable to
create N- terminated steps than it is to form Ti-terminated ones.
Since the chemical potential is not directly controlled or
measured in experiments, we are cautious about interpreting the
minus sign of the formation energy of the N-terminated steps, as
that sign can change with the choice of $\mu_N$. We can, however,
report that for a range of the chemical potential, $-8$ eV $\leq
\mu_N \leq -5$ eV, the relative formation energy
$\Lambda_{Ti}-\Lambda_{N}$ of the two types of edge terminations
is positive,  and lies in the range 689 meV/\AA $ <
\Lambda_{Ti}-\Lambda_{N} < $ 1690 meV/\AA. This reinforces the
conclusion that the N-terminated steps have a much lower formation
energy, and are therefore expected to be present in experiments
such as the ones described in \cite{K-SS-2002}. The physical
reason for the large difference in the formation energies of the
two types of [110] steps can be understood qualitatively by
considering the nature of the chemical bonding in TiN. In bulk,
the N atoms bond only to their nearest neighbors, i.e. six Ti
atoms. The titanium atoms bond not only with their six first-order
neighbors, but also with their second-order neighbors, i.e. 12 Ti
atoms. Therefore, while creating an N-terminated [110] step
implies breaking only three bonds N-Ti bonds, the formation of a
Ti step requires breaking of an additional seven Ti-Ti bonds.
Although some directional dependence of the strength of the
remaining Ti-Ti bonds may arise due to the presence of the step,
the broken Ti-Ti bonds will contribute to an increase in the
formation energy of the Ti steps. This effect is not counteracted
by atomic relaxations, which are comparable for the two types of
[110] steps.

\begin{figure}
\begin{center}
\epsfig{file=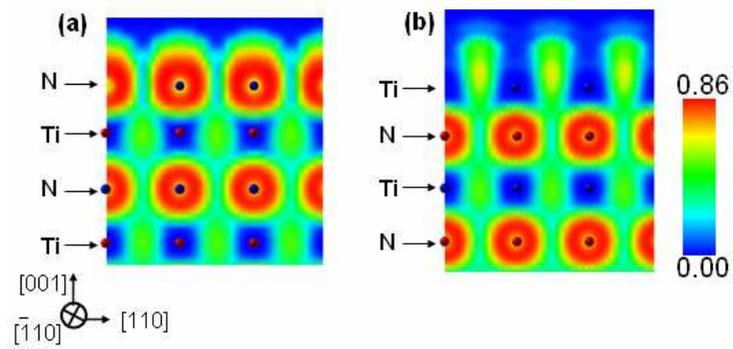,width=10.0cm} %
\caption{Electron localization function (arbitrary units) in an
(110) plane passing through a [110]-oriented step. In this plane,
the nature of the atomic rows alternates starting with N atoms on
the top row (step edge) (a), or with Ti atoms (b).
}%
\label{elfcars}
\end{center}
\end{figure}

The N atoms on the [110] step edge undergo rather large
displacements with respect to their bulk-truncated positions. As
shown in Fig.~\ref{displacements}(b), the edge moves inwards by
about 0.27\AA. This value is larger than the rumpling amplitudes
(refer to Table~\ref{flatdispl}) of the flat TiN(001), so it is
conceivable that the relaxation of the step can be observed
experimentally. For the Ti-terminated steps we also calculated
inward horizontal (vertical) relaxations of 0.2\AA \ ($\sim 0.1$
\AA). Since the horizontal displacements of the Ti steps are
comparable to those of the N-terminated steps, it is unlikely that
measurements of displacements can help identify the nature of the
atoms on the step edges. A better way of identifying the step
termination could perhaps be STM, since the electron localization
is different for the two types of steps. As illustrated in
Fig.~\ref{elfcars}, in the case of N-terminated steps the electron
density has maxima directly above the positions of the N atoms,
while for the Ti-steps these maxima lie between the Ti edge atoms.
Such local maxima of the electron localization function between
the Ti atoms occur not only on the surface, but also in the bulk
(Fig.~\ref{elfcars}): this indicates the presence of covalent
Ti-Ti bonds, which supports the above bond counting arguments for
the larger formation energy of the Ti steps.

In summary, we have studied the structure and energetics of steps
on TiN(001) surface using density functional calculations. For the
steps parallel to [100], we have obtained quantitative agreement
with recent experimental reports \cite{K-SS-2002}.  While a direct
comparison with the experiments was not possible in the case of
[110] steps, we have argued that the step edges are N terminated
-- a prediction that can be experimentally verified. We have also
reported the structures of the steps, and showed a rumpled
configuration for the [100] steps, and large inwards relaxations
for the [110] step edges. Future work aimed at determining the
kink formation energies on these steps can allow for the
calculation of their stiffness and for comparison with
experimental estimates.

 {\bf Acknowledgements:} We gratefully acknowledge
research support from NSF through the Brown University MRSEC
program (DMR-0079964) and Grants no. CMS-0093714 and CMS-0210095.
Computational support for this work was provided by the NCSA
(Grants no. MSS-030006 and DMR-020032N) and by the Graduate School
at Brown University through the Salomon Research Award. CVC thanks
Dr. S. Kodambaka for useful discussions about experiments
\cite{K-SS-2002,K-PRL-2002decay, K-PRL-2002,K-PRB-2003}.

\newpage
\newpage
\newpage

\newpage
\begin{table}
\begin{tabular}{l |cccccc}
\hline  \hline
  & $d_N ^1$ & $d_{Ti}^1$ & $d_N^2$ & $d_{Ti}^2$ & $r_1$ & $r_2$
  \\ \hline
Ref. \cite{Kobayashi}   & 0.134  & -0.074  &    &    & 0.21 & 0.005 \\
Ref. \cite{ukTiN-2000}  &   &   &  &  & 0.179 &  \\
Ref. \cite{uiucTiN-2003} & 0.12  & -0.06  &   &   & 0.18 &   \\
This work            & 0.113 & -0.066 & 0.040 & 0.028 & 0.179 & 0.012  \\
\hline \hline
\end{tabular}
\caption{Rumpling relaxation of TiN(001) compared with previous
reports. The outward displacements of the first ($d_{N, Ti}^1$)
and second($d^2_{N,Ti}$) layer surface atoms, and the rumpling
amplitudes $r_{1(2)}  \equiv d_N^{1(2)}-d_{Ti}^{1(2)}$ are given
in \AA .} \label{flatdispl}
\end{table}
\newpage
\begin{table}
\begin{tabular}{l l l | c l}
\hline  \hline
             &       &   & Present work   & Previous reports \\
$[100]$ steps: & $\Lambda_{[100]}$ &  (meV/\AA)  & 238 & 250$\pm 50$ \ \ \cite{K-SS-2002} \\
            &  $G_{[100]}$ & (meV\AA)             & 548 & ---  \\
$[110]$  steps:&  $\overline{\Lambda}_{[110]}$ & (meV/\AA) & 410  & 210$\pm 50$ \ \ \cite{K-SS-2002} \\
            &  $\overline{G} _{[110]}$ & (meV\AA)        & 445 &  --- \\
surfaces:   & $\gamma_{(001)}$  & (meV/\AA$^2$)          & 80.6  &81 \ \ \cite{ukTiN-2000,uiucTiN-2003} \\
            & $\overline{\gamma}_{(111)}$ & (meV/\AA$^2$) & 227.4 & 216 \ \ \cite{uiucTiN-2003} \\
\hline \hline
\end{tabular}
\caption{Step formation energies and interaction strengths
obtained from ab initio density functional calculations. Absolute
formation energies have been recently determined from experiments
by Kodambaka {\em et al. }\cite{K-SS-2002}. Surface energies of
TiN(001) and TiN(111) (averaged over N- and Ti- terminated
surfaces) are also given for comparison with previous
first-principles studies \cite{ukTiN-2000,uiucTiN-2003}. }
\label{beta-and-gamma}
\end{table}

\begin{thebibliography}{99}

% useful numbers
\bibitem{Kobayashi}K.~Kobayashi, Jpn. J. Appl. Phys. {\bf 39}, 4311
(2000); {\em ibid.} Surf. Sci. {\bf 493}, 665 (2001).

\bibitem{ukTiN-2000}M.~Marlo and V.~Milman,
Phys. Rev. B {\bf 62}, 2899 (2000).

\bibitem{uiucTiN-2003}D.~Gall, S.~Kodambaka, M.A.~Wall, I.~Petrov, J.E.~Greene,
J. Appl. Phys. {\bf 93}, 9086 (2003).


% Absolute orienatation dependent step energies on TiN(001)
\bibitem{K-SS-2002}S.~Kodambaka, S.V.~Khare, V.~Petrova, A.~Vailionis, I.~Petrov,
J.E.~Greene, Surf. Sci. {\bf 513}, 468 (2002).

% Decay kinetics of 2-d islands on TiN(111)
\bibitem{K-PRL-2002decay}S.~Kodambaka, V.~Petrova, S.V.~Khare, D.~Gall, A.~Rocket,
I.~Petrov, J.E.~Greene, Phys. Rev. Lett. {\bf 89}, 176102 (2002).

% Step energies from anisotropic island shape fluc
\bibitem{K-PRL-2002}S.~Kodambaka, V.~Petrova, S.V.~Khare,
D.D.~Johson, I.~Petrov, J.E.~Greene, Phys. Rev. Lett. {\bf 88},
146101 (2002).


%
% Absolute-orientation-dependent step formation energies on TiN(111)
%
\bibitem{K-PRB-2003}S.~Kodambaka, S.V.~Khare, V.~Petrova,
D.D.~Johnson, I.~Petrov, J.E.~Greene, Phys. Rev. B {\bf 67},
035409 (2003).

% Wulff construction Khare paper, absolute orientation dependent step formation energies
%\bibitem{khare}S.V.~Khare, S.~Kodambaka, D.D.~Johnson, I.~Petrov,
%J.E.~Greene, Surf. Sci. {\bf 522}, 75 (2003).

% VASP references

\bibitem{Vasp}G.~Kresse and J.~Furthmuller, Phys. Rev. B {\bf 54}, 11169
(1996); {\em ibid.}, Comput. Mater. Sci. {\bf 6}, 15 (1996).

%\bibitem{paw}G.~Kresse, and J.~Joubert, Phys. Rev. B {\bf 59}, 1758 (1999).

\bibitem{pw91}J.~Perdew and Y.~Wang, Phys. Rev. B {\bf 45}, 13244 (1992).

\bibitem{Poon}T.W.~Poon, S.~Yip, P.S.~Ho and F.F.~Abraham, Phys.
Phys. Rev. B {\bf 45}, 3521 (1992).

%prior paper on TaC
\bibitem{tacpaper}V.B.~Shenoy and C.V.~Ciobanu, Phys. Rev. B {\bf
67}, 081402(R) (2003).

% dipolar repulsion

\bibitem{marchenko}V.~I.~Marchenko and Y.~A.~Parshin, Sov. Phys. JETP
{\bf 52}, 129 (1980).

\bibitem{Jay}C.~Jayaprakash, C.~Rottman and W.~F.~Saam, Phys.
Rev. B {\bf 30}, 6549 (1984).

\end{thebibliography}
\end{document}